\documentclass{article}
\usepackage[T1]{fontenc}
\usepackage[utf8]{inputenc}
\usepackage{iclr2026_conference,times}

\usepackage{amsmath,amsfonts,bm}

\def\eqref#1{equation~\ref{#1}}
\def\1{\bm{1}}

\DeclareMathAlphabet{\mathsfit}{\encodingdefault}{\sfdefault}{m}{sl}
\SetMathAlphabet{\mathsfit}{bold}{\encodingdefault}{\sfdefault}{bx}{n}

\usepackage{hyperref}
\usepackage{url}
\usepackage{graphicx}
\usepackage{subcaption}
\usepackage{booktabs}
\usepackage{microtype}

\title{Dreaming of Others: Latent Teammate Modeling in World Models\\ for Multi-Agent Reinforcement Learning}

\author{Tomas Leroy-Stone
}

\iclrfinalcopy % Uncomment for camera-ready version
\begin{document}
\maketitle

\begin{abstract}
In cooperative multi-agent reinforcement learning (MARL), agents must coordinate with partners whose internal policies and intentions are not directly observable. While world models such as Dreamer have demonstrated strong generalization and sample efficiency in single-agent settings, their application to MARL remains limited by an inability to handle teammate-induced uncertainty. We propose a new perspective: treat teammates as structured, learnable components within the agent's world model. We introduce an architecture that factorizes the latent state of a Dreamer-style recurrent state-space model (RSSM) into environment and teammate components, and learns an auxiliary Theory-of-Mind (ToM) head to infer latent embeddings of partner behavior such as character, intent, and predicted actions from partial trajectories. These teammate latents condition the actor and critic, enabling the agent to imagine and adapt to diverse collaborators. We outline how this approach can support zero-shot and few-shot coordination in partially observable settings and propose a set of benchmarks and evaluation protocols to assess its impact. This work positions world models as not only predictors of environmental dynamics, but as simulators of social behavior, opening new directions for generalizable, human-compatible AI.
\end{abstract}

\section{Introduction}
World models have emerged as a powerful paradigm for reinforcement learning (RL), enabling agents to learn compact latent dynamics and train decision policies from imagined trajectories \citep{hafner2025mastering}. Despite their success in single-agent domains, deploying world models in cooperative multi-agent RL (MARL) remains challenging because the behavior of teammates introduces non-stationary, partially observable structure that is typically compressed as undifferentiated noise.

We advocate extending world models from simulating environments to simulating \emph{others} within them. Concretely, we propose a teammate-conditioned world model that (i) factorizes the latent state into environment and teammate components and (ii) augments the model with a Theory-of-Mind (ToM) head that infers latent embeddings of partner behavior from partial histories. The resulting teammate latents condition the actor and critic during imagination, allowing agents to anticipate partner variability and adapt to unseen collaborators without access to their internal observations or policies.

This focus on collaboration with humans removes many of the simplifications common in multi-agent reinforcement learning. Agents cannot rely on centralized training with privileged information, explicit communication channels, or access to a partner’s policy, goals, or observations. The human partner instead appears as a partially observed, adaptive process whose behavior must be inferred and integrated into the agent’s predictive model. 

This paper is a proposal intended to catalyze discussion in the world-models community; we do not report empirical results. We precisely describe the architecture, objectives, and evaluation protocol, argue why modeling teammates as structured latent processes can reduce apparent non-stationarity, and outline how to assess zero-shot and few-shot coordination. The broader goal is human AI collaboration, where partners are partially observable and heterogeneous \citep{carroll2019utility,liang2024learning}.

\section{Background and Related Work}

\paragraph{World Models and Dreamer.}
World models learn a generative model of environment dynamics to support prediction, planning, and control in latent space. By compressing high-dimensional observations into a recurrent latent representation, these models allow agents to reason about abstract dynamics and train policies through imagination rather than direct interaction. The Dreamer family \citep{hafner2025mastering} exemplifies this paradigm: it combines a Recurrent State-Space Model (RSSM) that captures both deterministic memory and stochastic latent variables with an actor-critic trained entirely through latent imagination rollouts. The approach has demonstrated remarkable generalization and sample efficiency, mastering hundreds of tasks across diverse continuous-control domains with fixed hyperparameters. DreamerV3 in particular shows that large world models can scale and generalize to unseen tasks without task-specific tuning, suggesting that latent imagination provides a powerful inductive bias for learning structured, reusable dynamics. 

\paragraph{Cooperative Multi-Agent Reinforcement Learning.}
Cooperative MARL studies how multiple agents can learn to coordinate toward a shared reward function under partial observability. The most common training paradigm, Centralized Training with Decentralized Execution (CTDE), provides agents with access to joint information during training while requiring independent execution policies at test time. Policy-gradient algorithms such as MAPPO \citep{yu2022surprising} and value-decomposition methods such as QMIX \citep{rashid2020monotonic} have become standard baselines, offering empirical robustness and scalability across benchmark tasks. In these settings, non-stationarity remains a central challenge: as teammates update their policies, the effective environment observed by each agent changes.

\paragraph{World Models for MARL.}
Recent works have begun to extend world models to multi-agent coordination, highlighting the promise and limitations of imagination-based approaches in social environments. MA-Dreamer \citep{lobos2022ma} introduces shared imagination between agents, enabling communication through latent rollouts. CoDreamer \citep{toledo2024codreamer} decentralizes this process, using latent message passing to synchronize multiple local world models. GAWM \citep{shi2025gawm} proposes a global-aware latent representation to capture collective state information across agents, while MACD explores counterfactual imagination to improve credit assignment \citep{chai2024aligning,egorov2022scalable}. Although these studies advance model-based cooperation, they often assume shared access to latent states or centralized imagination, effectively sidestepping the challenge of partial observability. 

\paragraph{Theory of Mind and Agent Modeling.}
Parallel research in agent modeling and Theory of Mind (ToM) explores how artificial agents can infer the beliefs, goals, or strategies of others. ToMnet \citep{rabinowitz2018machine} introduced the idea of learning to predict another agent’s future actions from past trajectories, laying the foundation for meta-learning and inference-based models of social reasoning. Subsequent work formalized intent inference under partial observability \citep{papoudakis2021agent}, demonstrating improved robustness and interpretability in cooperative and competitive settings. In human-AI coordination, explicitly modeling partner preferences and intentions improves adaptability and alignment \citep{carroll2019utility,liang2024learning}. 

\section{Teammate-Conditioned World Models}
\subsection{Design Rationale}
In cooperative settings, observations reflect both physical dynamics and the evolving behavior of teammates. Treating other agents as exogenous noise induces apparent non-stationarity as partners change policies. We posit that modeling teammates as structured latent processes, separate from environmental latents, can reduce non-stationarity, produce socially consistent imagined rollouts, and enable rapid adaptation to new partners.

\subsection{Architecture}
At each time step $t$, the encoder consumes the controlled agent's observation $x_t$ and action $a_t^0$ to produce a deterministic hidden state $h_t$. We posit a factorized stochastic latent
\[
z_t = \big[z_t^{\text{env}},\; z_t^{\text{team}}\big],
\]
where $z_t^{\text{env}}$ captures environment dynamics and $z_t^{\text{team}}$ encodes inferred teammate behavior.

Two decoders operate on these latents. An observation decoder maps $z_t^{\text{env}}$ to $\hat{x}_t$. A teammate-policy decoder maps $z_t^{\text{team}}$ to a predictive distribution $\hat{\pi}_t^j(\cdot)$ over the teammate's next action. Actions from all agents $(a_t^0, a_t^j)$ drive the transition to $h_{t+1}$. During actor-critic learning, the hidden state and teammate latents condition policy and value heads to generate $(a_t^0, v_t, r_t)$ and support imagination with sampled $z_t^{\text{team}}$.

We retain standard Dreamer training for the world model and control components \citep{hafner2025mastering} and add a dedicated ToM objective for teammate modeling, without requiring shared imagination, centralized policies, or explicit communication \citep{lobos2022ma,toledo2024codreamer,shi2025gawm}.

\begin{figure}[t]
    \centering
    \includegraphics[width=0.9\linewidth]{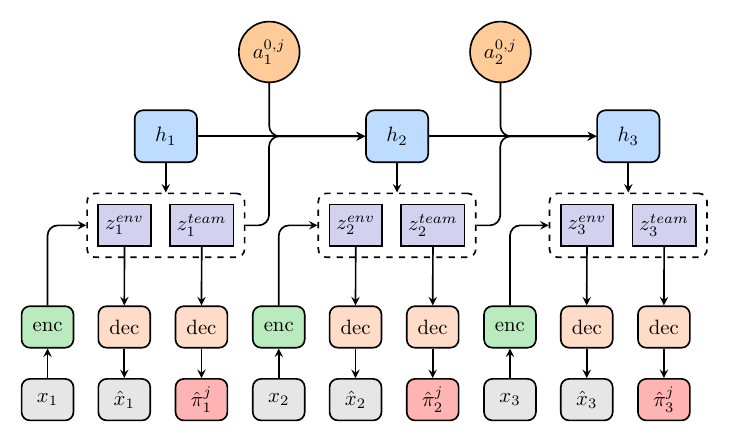}
    \caption{\textbf{World model and teammate modeling.} An RSSM with factorized latent $z_t=[z_t^{env},z_t^{team}]$. The decoder reconstructs $\hat{x}_t$ from $z_t^{env}$ and predicts teammate policy $\hat{\pi}_t^j(\cdot)$ from $z_t^{team}$. Actions $(a_t^0,a_t^j)$ update the transition to $h_{t+1}$. The ToM loss supervises $\hat{\pi}_t^j$.}
    \label{fig:worldmodel}
\end{figure}

\subsection{Teammate-Modeling Objective}
Let $\pi_t^j(\cdot)$ denote the empirical or behavior distribution of the teammate's next action at time $t$ obtained from data, and $\hat{\pi}_t^j(\cdot)$ the model's prediction from $z_t^{\text{team}}$. We minimize a calibrated cross-entropy with temporal regularization:
\begin{equation}
\label{eq:tomloss}
\mathcal{L}_{\text{ToM}}
=
\mathbb{E}_{t}\Big[
-\sum_{a}\pi_t^j(a)\,\log \hat{\pi}_t^j(a)
\Big]
+
\alpha\,\mathrm{KL}\!\left(
q\!\left(z_t^{\text{team}}\mid h_t\right)\,\middle\|\,p\!\left(z_t^{\text{team}}\mid h_{t-1},a_{t-1}\right)
\right),
\end{equation}
where the KL term stabilizes temporal consistency of the inferred teammate latent. In practice, $\pi_t^j$ can be a one hot target from the observed $a_t^j$ or a smoothed label. The total objective augments standard Dreamer training with $\lambda_{\text{ToM}}\mathcal{L}_{\text{ToM}}$.

\begin{figure}[t]
    \centering
    \includegraphics[width=0.8\linewidth]{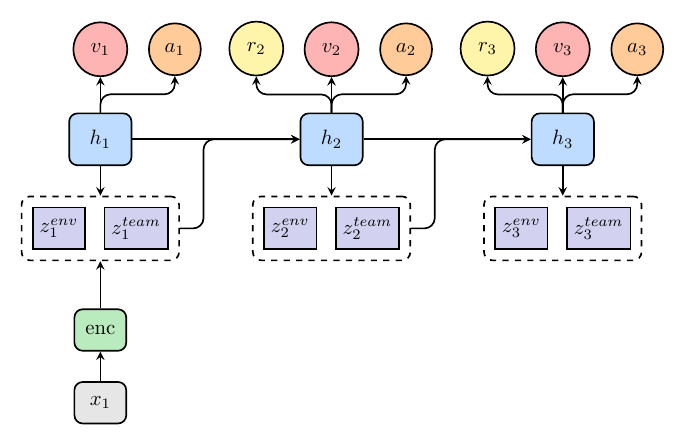}
    \caption{\textbf{Actor-critic imagination.} The hidden state and teammate latents condition the policy and value heads to produce $(a_t^0, v_t, r_t)$. Imagination samples $z_t^{\text{team}}$ to simulate partner variability for zero-shot and few-shot coordination.}
    \label{fig:actorcritic}
\end{figure}

\subsection{Deployment and Adaptation}
At test time, the model infers $z_t^{\text{team}}$ online from observed teammate actions and conditions the actor and critic on this embedding. Imagined rollouts sample plausible teammate trajectories, enabling zero-shot coordination with unseen partners and few-shot improvement as more behavior is observed, without access to the partner's policy or observations.

\section{Evaluation Protocol}
We propose an evaluation protocol that emphasizes generalization and adaptability across partners and settings. Multi-Agent Particle Environments \citep{lowe2017multi} provide lightweight, controllable scenarios for diagnosing whether the model learns identifiable teammate latents and separates social from physical dynamics. Overcooked-AI \citep{carroll2019utility} serves as the main testbed for zero-shot coordination, using standard partner splits to assess how well the inferred teammate embedding supports immediate coordination with unseen collaborators. Melting Pot \citep{leibo2021scalable} evaluates robustness across diverse social contexts and partner populations at scale, probing how the method copes with distributional shift in norms and conventions. Across all environments, we plan to leverage unified tooling from JaxMARL \citep{rutherford2024jaxmarl} and BenchMARL \citep{bettini2024benchmarl}, which make significant strides toward standardizing environments, metrics, and baselines. We will report episodic return and sample efficiency, a zero-shot coordination score with unseen partners, few-shot improvement after a small number of interactions and cross-play robustness across partner pairings.

\section{Discussion and Outlook}
We present a conceptual architecture that integrates teammate modeling into world models for MARL. By treating other agents as structured latent processes rather than stochastic noise, the proposed design provides a path toward socially grounded world models. This formulation offers several potential benefits: reduced non-stationarity, improved zero-shot coordination, interpretable latent representations, and greater compatibility with human partners.

Ultimately, teammate-conditioned world models may help agents not only dream of the worlds they inhabit, but also of the minds that share them.

\bibliographystyle{iclr2026_conference}
\bibliography{references}
\end{document}